\documentclass[fleqn,10pt]{wlscirep}
\title{Optogenetically Induced Spatiotemporal Gamma Oscillations in Visual Cortex}

\author[1]{Fereshteh Arab}
\author[2, 5]{Sareh Rostami}
\author[3, 5]{Mohammad Dehghani-Habibabadi}
\author[1, 4 *]{Vahid Salari}
\author[2,*]{Mir-Shahram Safari}

\affil[1] {Department of Physics, Isfahan University of Technology, Isfahan 84156-83111, Iran,}
\affil[2, **] {Neuroscience Research Center, Shahid Beheshti University of Medical Sciences, Tehran, 19615-1178, Iran,}
\affil[3, **] {Institute for Theoretical Physics, University of Bremen, 28359 Bremen, Germany,}
\affil[4]{Department of Physical Chemistry, University of the Basque Country (EPV/EHU), Bilbao, Spain.}
\affil[5, **]{Equal contribution}

\affil[1*]{vahid.salari@ehu.eus}
\affil[2*]{safari@sbmu.ac.ir}

\begin{abstract}

It has been hypothesized that Gamma cortical oscillations play important roles in numerous cognitive processes and may involve psychiatric conditions including anxiety, schizophrenia, and autism. Gamma rhythms are commonly observed in many brain regions during both waking and sleep states, yet their functions and mechanisms remain a matter of debate. Spatiotemporal Gamma oscillations can explain neuronal representation, computation, and the shaping of communication among cortical neurons, even neurological and neuropsychiatric disorders in neo-cortex. In this study, the neural network dynamics and spatiotemporal behavior in the cerebral cortex are examined during Gamma brain activity. We have directly observed the Gamma oscillations on visual processing as spatiotemporal waves induced by targeted optogenetics stimulation. We have experimentally demonstrated the constant optogenetics stimulation based on the ChR2 opsin under the control of the CaMKII$\alpha$ promotor, which can induce sustained narrowband Gamma oscillations in the visual cortex of rats during their comatose states. 
The injections of the viral vector [LentiVirus – CaMKII$\alpha$ – ChR2] was performed at two different depths, 200 and 500 $\mu$m. Finally, we computationally analyze our results via Wilson-Cowan model.

\end{abstract}

\begin{document}

\flushbottom
\maketitle
\thispagestyle{empty}

\section*{Introduction}

One of the powerful experimental approaches in neuroscience to understand the spatiotemporal neuronal network is optogenetics.
According to many types of research, optogenetics is a technique that involves using light to control cells in live tissues, typically neurons, that have been genetically modified to express light-sensitive ion channels.
Optogenetics is a neuro-modulation method that uses a compound of techniques from optics and genetics to control and monitor individual neural activities in living tissues and precisely measure these manipulation effects in real-time, \cite{deisseroth2006next, marchhistory, yizhar2011optogenetics}, and it enables us to manipulate neuronal activities in a millisecond. Optogenetics has furthered the fundamental scientific understanding of how specific cell types contribute to the function of biological tissues such as neuronal circuits in-vivo. Moreover, on the clinical side, optogenetic-driven research has led to insights into Parkinson's disease and other neurological and psychiatric disorders. Indeed, optogenetics has also provided different insights into neuronal codes relevant to Autism, Schizophrenia, drug abuse, anxiety, and depression. \cite{deisseroth2011optogenetics,fenno2011development,yizharbest,deisseroth2012optogenetics}. Cortical oscillations are rhythmic patterns of neuronal activity that are synchronized across many neurons within or across brain regions that are classified based on frequency into Delta (0.5-4 Hz), Theta (4-8 Hz), Alpha (8-12 Hz), Beta (13-30 Hz), and Gamma (30-100 Hz) oscillations. \cite{hari1997human,buzsaki2004neuronal}
 Cortical oscillation in Theta and Gamma frequency range has been hypothesized to play essential roles in numerous cognitive processes and may involve psychiatric conditions including anxiety, schizophrenia, and autism. Gamma rhythms are commonly observed in many brain regions during both waking and sleep states, yet their functions and mechanisms remain a matter of debate \cite{}.

\subsection*{History}
Spatiotemporal Gamma oscillations can explain neuronal representation, computation, and the shaping of communication among cortical neurons, even neurological and neuropsychiatric disorders in neo-cortex.\cite{buzsaki2012mechanisms,tiesinga2009cortical,ray2015Gamma,womelsdorf2006Gamma,van2014alpha}
Recent studies have used optogenetic stimulation to probe the mechanisms and neural circuits underlying Gamma oscillations' generation in the neo-cortex. \cite{sohal2012insights,csicsvari2003mechanisms,traub1996analysis,yamamoto2014successful, buhl1998cholinergic}. Moreover, different investigations have been done on optogenetically induced Gamma oscillations, such as compared Gamma oscillations in entorhinal-hippocampal system \cite{butler2018comparison}, neuroprotective effects due to chronic gamma entrainment \cite{adaikkan2020gamma}, and a mechanistic framework for modulation of pain by specific activity patterns in the primary somatosensory cortex \cite{tan2019gamma}. In a recent optogenetic study, it has been elucidated the substrate oscillations in the prefrontal cortex via a protocol with a combination of optogenetics and electrophysiological recordings from neonatal mice, where the light was transferred to the layer II/III pyramidal neurons by in utero electroporation with Channelrhodopsin, boosting network oscillations within Gamma-Beta frequency range \cite{bitzenhofer2017layer}. However, the activation of layer V/VI pyramidal neurons caused nonspecific network activation \cite{bitzenhofer2017layer}. In an investigation \cite{butler2016intrinsic}, it was expressed channelrhodopsin-2 under the CaMKII$\alpha$ promoter in mice and demonstrates that the cornuammonis area 1 (CA1) is capable of generating intrinsic Gamma oscillations in response to Theta input. This Gamma generator is independent of activity in the upstream regions, highlighting that CA1 can produce its own Gamma oscillation in addition to inheriting activity from the upstream regions, which supports the theory that Gamma oscillations predominantly function to achieve local synchrony and that a local Gamma generated in each area conducts the signal to the downstream region. Additionally, in another study, the optogenetic stimulation was performed on two macaque monkeys during their awake resting and reach and grasp states in the primary motor (M1) and ventral premotor (PMv) cortices of two subjects. They used the particular optogenetic construct includeing red-shifted opsin C1V1 under the control of the CaMKII$\alpha$ promoter and AAV5 viral vector. 

\subsection*{This study}
In this paper, the effect of induced Gamma oscillations on visual power is investigated, and we check whether the visual activity of subjects increases these oscillations in the cortex or not. We would like to probe the connection between visual power and the power of induced Gamma oscillations in the second and the fifth layers of cortical. We investigate optogenetically induced Beta and Gamma oscillations in rat's visual cortex by Adeno Associated Virus (AAV) method. We express the channelrhodopsin2 in layers II and V, simultaneously with visual stimulation. As a novel aspect of our work, we have examined whether the visual response is affected by the optogenetic laser or not. In fact, the present study examines and analyzes the effect of induced Gamma oscillations by targeted optogenetics stimulation, recorded by intra-cortical one-channel extracellular techniques in rat's visual cortex during their comatose states. Injections of the viral vector [ LentiVirus - ChR2 – CaMKII$\alpha$] were performed at two different depths ( 200 $\mu$m and 500 $\mu$m). During trial stimulation, a 470-nm blue laser was delivered light through the polymer optical fiber. LFP trials in anesthetized rats in 4 states are recorded.

\subsection*{Methodology} 
Three requirements are necessary to apply optogenetic techniques: selecting optogenetic molecular regents (OMRs), targeted expression of OMRs in the neurons or regions of interest, and a light delivery system\cite{yawo2013optogenetic}. Neuronal control can be achieved using optogenetic actuators like Channelrhodopsin. An engineered viral vector is used to introduce the microbial opsin to a specific region of the organism, containing the optogenetics actuator gene attached to a recognizable promoter CaMKII$\alpha$. \cite{zhang2008red,nagel2002channelrhodopsin, berthold2008channelrhodopsin,mattis2012principles,yizhar2011optogenetics,zhang2011microbial}. Various methods have been devised to express the OMRs in the cells to be controlled specifically, such as an adeno-associated virus (AAv) or lentivirus incorporating a target cell-specific promoter as a vector. \cite{zhang2006channelrhodopsin,fenno2011development,gradinaru2010molecular}
In this study, lentivirus was used. The changes in membrane potential, which show dynamic fluctuations in response to excitatory and inhibitory inputs, play key roles in excitable cells such as neurons and muscle cells. Electrophysiological approaches, including the micro-pipette method and the patch-clamp methods, allow recording fast voltage changes at a sub-millisecond level (e.g., action potentials), i.e. used here. \cite{scanziani2009electrophysiology,ting2014acute,wang2015optogenetics}\\

 From theoretical point of view, we model the cortex as a Wilson-Cowan neural field \cite{heitmann2017optogenetic}. We consider laser stimulus as a source of an external current into the neuronal populations. During the laser onset, the outward current is applied to the neuronal population. We adjust the model parameters to explain our experimental results for describing variations in Beta and Gamma oscillations power.
 
 \subsection*{Computational Data Analysis} 

\begin{figure}[h]
\centering
\includegraphics[width=9cm, height=7cm]{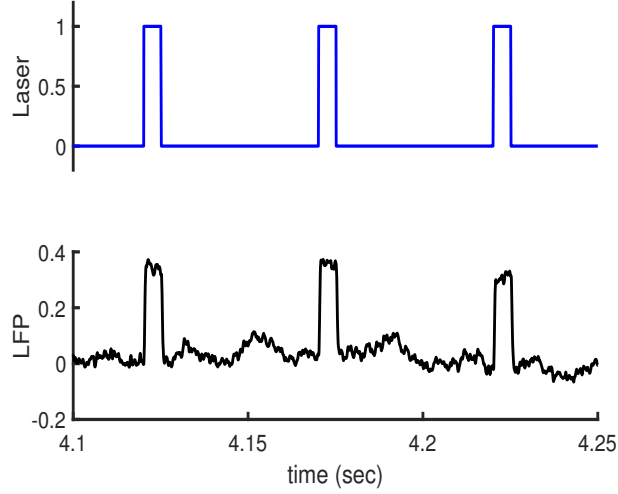}
\caption{(Color online) Blue line shows amplitude of laser in time. Red line is corresponding LFP during laser stimulus. }
\label{fig:f5}
\end{figure}

In this research, we use FIR1 filter with a band-pass type. First, we filter data in the range of 5 - 300 Hz, and then we focus on the range of Beta and Gamma frequencies. One of the quantities that obtain differences in four different conditions (Control, Visual, Laser, and Laser+Visual) is the mean power signal. Signal power in Gamma range from 30 - 150 Hz and Beta range 13 - 30 Hz were calculated for different trials. Equation~\ref{eq1} is the formula of discrete-time signal \textit{x(n)} power.
\begin{equation} \label{eq1}
\begin{split}
{\text{Signal Power}} = \frac{1}{2N-1}\sum_{n=-N}^{n=N}\left  |x(n) \right |^{2}
\end{split}
\end{equation}
To find the pure effect of Optogenetic laser at different depths, we define equation~\ref{eq2}, as a biological response to laser stimulus.
\begin{equation} \label{eq2}
\begin{split}
{\text{Biological Response}} = (P_{\text{Laser+Visual}}-P_{\text{Laser}})-(P_{\text{Visual}}-P_{\text{Control}})
\end{split}
\end{equation}
where P is the power of the signal. P$_{\text{Visual}}$ - P$_{\text{Control}}$ is any changes in the visual trials' power compared to the control trials. If the quantity gets positive, it means that the visual stimulus has a positive effect in Beta band frequencies.\\
P$_{\text{Visual}}$+$_{\text{Laser}}$ - P$_{\text{Laser}}$ shows changes caused by vision. If Biological Response is equal to zero, there is no effect from optogenetic stimulus on visual responses.\\
In this simulation, whenever the Laser is on, the corresponding LFPs data are not considered. We concentrate on the biological response between two pulses.
As figure~\ref{fig:f5} shows, blue lines are non-zero at the time of Laser is switched on, and the corresponding LFP is shown in red color at the same time.
Figure~\ref{fig:f5}, shows at the time of Laser is switched on, there is a jump in LFP data, which gives a high frequency in the analysis. By removing these data and focusing on time between two pulses, artifacts do not produce high frequencies.

\subsection*{Modeling Approach}
\begin{figure}[h]
\centering
\includegraphics[width=7cm, height=5cm]{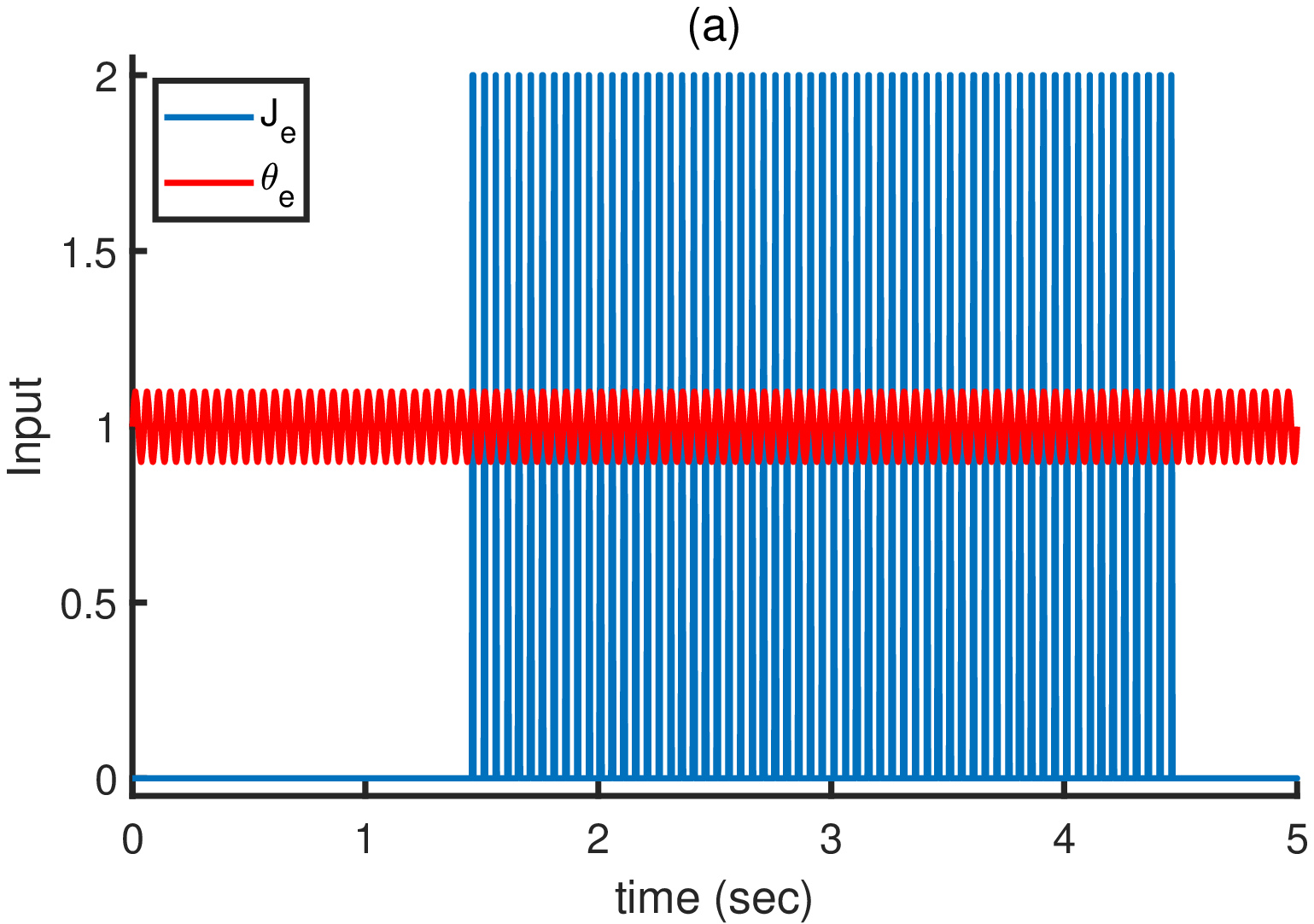}
\includegraphics[width=7cm, height=5cm]{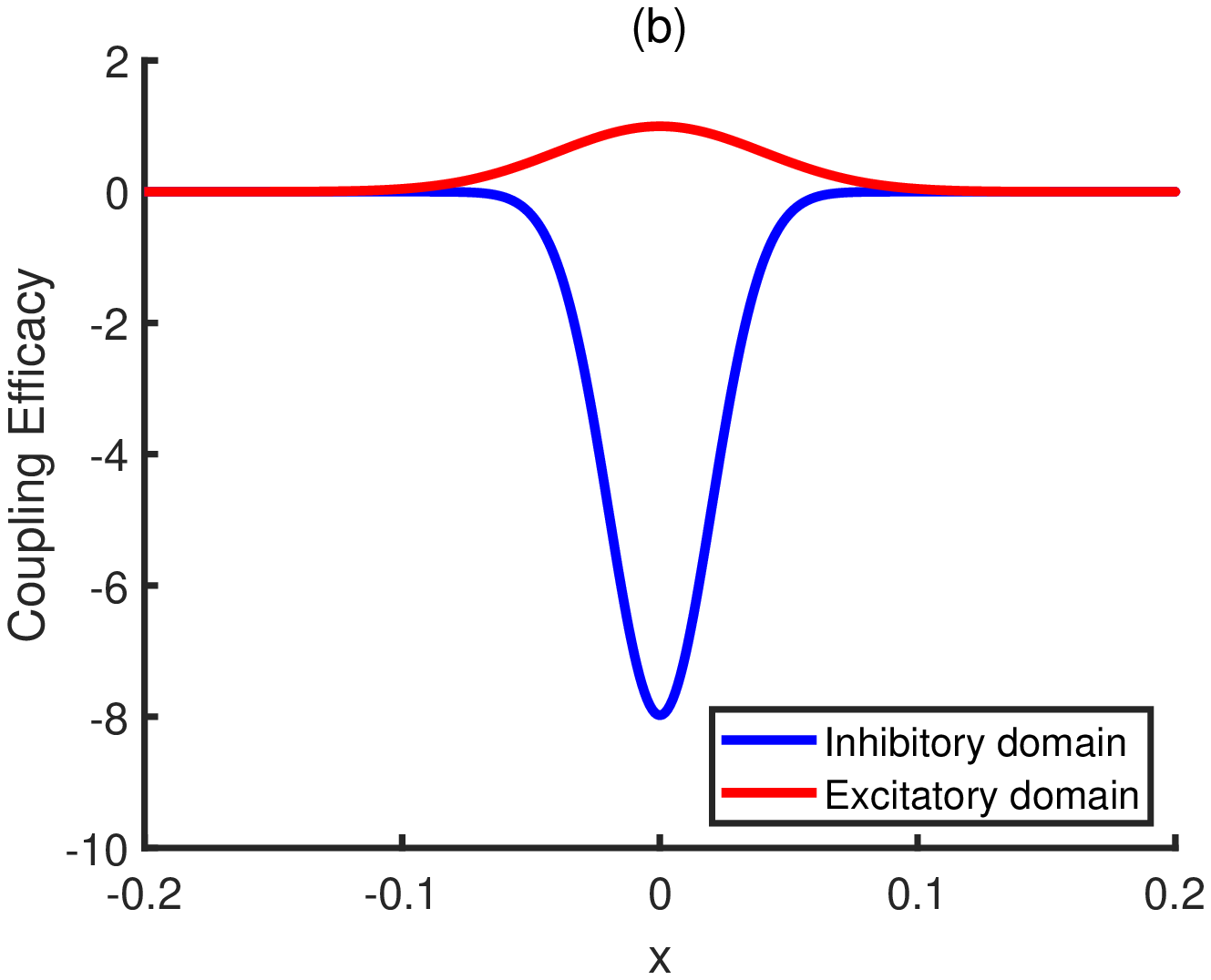}
\caption{(color online) Left: Two different types of inputs to the population. Right: Excitatory and inhibitory population domain and efficacy.}
\label{fig:f6}
\end{figure}

We use the Wilson Cowan model (as a standard model in optogenetics) \cite{wilson1973mathematical, wilson1972excitatory} for modeling our study as follows in the equation ~\ref{eq3}.

\begin{equation} \label{eq3}
\begin{split}
\tau_{e}\frac{\mathrm{}dE(r,t) }{\mathrm{} dt}&= - E(r,t) + f(W_{ee}*E(r,t)+W_{ei}*I(r,t)+\theta _{e} +J _{e})\\
\tau_{i}\frac{\mathrm{}dI(r,t) }{\mathrm{} dt}&= - I(r,t) + f(W_{ie}*E(r,t)+W_{ii}*I(r,t)+\theta _{i}+J _{i})
\end{split}
\end{equation}
E and I represent mean firing rates of the excitatory and inhibitory population at the point r, (x,y), respectively. Each population has a specific membrane time constant $\tau$, and $\theta$ is a continuously synaptic current, visual stimulation, or other currents to the cells. The variables $J _{e}$ and $J _{i}$ are excitatory and inhibitory currents to each cell respectively, due to light in the optogenetics setup \cite{lu2015optogenetically}. The Gaussian distribution is considered for the domain and efficacy of inhibitory and excitatory synapses, and the parameter $W$ is defined based on this distribution, i.e. the synaptic influence of each population on itself and other cells,as follows
\begin{equation} \label{eq4}
\begin{split}
W=\frac{w}{\sqrt{2\pi s}}\exp(\frac{x^{2}}{2s^{2}})
\end{split}
\end{equation}
$w$ and $s$ depend on inhibitory and excitatory populations are different. Figure ~\ref{fig:f6} b shows inhibitory and excitatory cells power and effect's domain. There are four types of connections, and w is changed depending on the type of connection. inhibitory to excitatory, $w_{ie}$, inhibitory to inhibitory, $w_{ii}$, excitatory to excitatory, $w_{ie}$, excitatory to inhibitory, $w_{ie}$. In our study $w_{ii}$ and $w_{ie}$ is -0.4 and $w_{ee}$ and $w_{ie}$ is +0.1. $s_{i}$ is 0.04 and $s_{i}$ is 0.02.

\begin{equation} \label{eq5}
K_{ie}(x)*I(x,t)=\int K_{ie}(x-y)I(x,t)dy
\end{equation}

In this study, we have considered 100 inhibitory and 100 excitatory cells, and each cell dose not have connection to itself. The function f is the threshold function, which is written in equation~\ref{eq5}.
\begin{equation} \label{eq6}
\begin{split}
f(x)=\frac{1}{1+e^{-4(u-1)}}
\end{split}
\end{equation}
where u is local synaptic input at spatial position r at time t. LFP is a contribution of excitatory and inhibitory populations activity, equation ~\ref{eq6}.
\begin{equation} \label{eq7}
\begin{split}
\text{LFP}=0.8E + 0.2 I
\end{split}
\end{equation}

Generating oscillations depend on the inputs to the populations, and there are different inputs to the populations in different trials. We consider that there is no input to inhibitory cells, therefore, $\theta_{i} =0$ and $J_{i} =0$.  Figure ~\ref{fig:f6} a is shown different type of current to excitatory population.
A constant input plus a periodic Beta frequency,15 Hz, input \emph{$1+sin(15\pi t)$} to the excitatory population ensures Beta and Gamma oscillations in the control state. in the beginning $\tau_{i} = 4 ms$ and $\tau_{e} = 2 ms$  
We consider a periodic triangular laser stimulus with an absolute amplitude to the excitatory neurons, and laser time onset is the same as the experiment. Opsin's main effect is on excitatory neurons. Therefore laser stimulus is considered only for the excitatory cells.

\section*{Materials and Methods}\label{Materials and Methods}
The general setup of the experiment is shown in Figure~\ref{fig:f0}. The details of experiment are as follows:

\begin{figure}
\centering
\includegraphics[width=16cm, height=9cm]{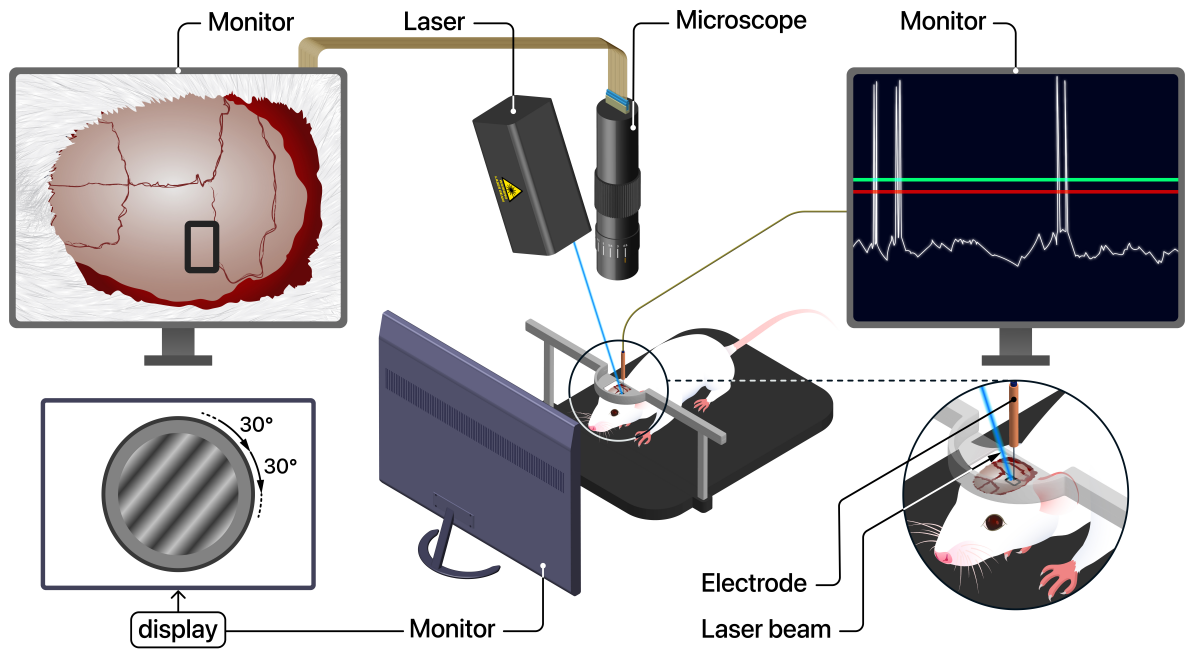}
\caption{Setup of the experiment. Here, in order to record V1 data in the left hemisphere, the monitor was located in the right side of animal instead of front side.}
\label{fig:f0}
\end{figure}

\subsection*{Experimental Task}
\subsubsection*{Animal preparations and virus injection}

Ten naive rats of either sex, weight 120 – 150 g, were kept under standard conditions at 20$\scriptstyle\pm$2 $^{\circ}$C and light: dark hours of 12:12 with free access to water and standard food. All experimental procedures were performed according to the Guide for the Care and Use of Laboratory Animals (National Institute of Health Publication No. 80-23, revised 1996) and were approved by the ethics committee, Shahid Beheshti University of Medical Sciences, ethical code IR.SBMU.PHNS.REC. 1398.024. Animals were anesthetized by ketamine/xylazine (ip, 80/2.5 mg/kg). Their body temperature maintained at 36.0 – 36.5$^{\circ}$C by a rectal thermos probe feeding back to a heating pad (ATC-402, Unique Medical). A small custom-made head plate to stabilize the animal‘s head under virus injection and LFP recording attached over the occipital region of the left hemisphere. After determining of cranial window over primary visual cortex (V1) (3-9 mm posterior to Bregma; 1-5 mm lateral), a small part of the skull and dura mater were removed and exposed cortex was covered by Ringer’s solution. We injected viral vector encoding Channelrhodopsin-2 under the control of CaMKII promotor (1.5- 2 $\mu$l) in layers II-IV (200 - 500 $\mu$m) of V1 by applying positive pressure to glass electrode (pipette) preparing for LFP recording in same area two weeks later.

\subsubsection*{Data acquisition and visual stimulation}

Square-wave gratings (0.08 cycle/deg) at 100 mW/mm$^2$ contrast was moved on an LCD monitor in 12 directions at 30 degree steps (0-360 degrees). These 12 patterns of visual stimuli (2 second before, 3 second during and 2 second after the presentation of each stimulus, respectively) were presented 3 times in a randomly shuffled order. The LCD monitor covered 80*50 degrees of the visual field at a viewing distance of 28 cm.

\subsubsection*{Optogenetic stimulation}

A blue laser light (470 nm) was directed into an optical fiber (700 $\mu$m diameter), mounted on a holder for positioning in the recording chamber. The laser's output function was tested with the optical power meter at the tip of the optical fiber. Only a low total output power may be needed to achieve ChR2 activation. We used three trials in different laser intensities (input: 50 mW/mm$^2$ and output: 5 mW/mm$^2$), input: 75 mW/mm$^2$ and output: 31 mW/mm$^2$), (input: 100 mW/mm$^2$ and output 80 mW/mm$^2$), respectively. Each trial consists 10 pulses at 20 Hz (2 s light off, 3 s light on (5 ms delay, 45 ms during)).

\subsubsection*{LFP recording}

Animal were anesthetized with Urethane (ip, 1.5 mg/g body weight) and then placed in a stereotaxic frame (craniotomy). LFP recording from layer II (200 $\mu$m) and layer IV (500 $\mu$m) of V1 were performed by recording electrodes pulled from borosilicate glass capillary with filaments (0.86 mm inner diameter, 1.5 mm outer diameter) filled by Ringer’s solution. The resistance of these electrodes was 500 k$\Omega$. LFPs were recorded as four groups (three trials for each): 1) without visual stimulation, 2) without visual stimulation but with optogenetic stimulation, 3) with visual stimulation, 4) with visual stimulation and optogenetic stimulation. Using a molecular device amplifier (Axopatch 200B, Molecular Devices), recorded data was sampled at 20 kHz, filtered at 2-5 KHz, digitized at 20 KHz, and fed into a personal computer with an NI-DAQ board (PCI-MIO-16E-4, National Instruments).


\section*{Results}

\begin{figure}
\centering
\includegraphics[width=9cm, height=7cm]{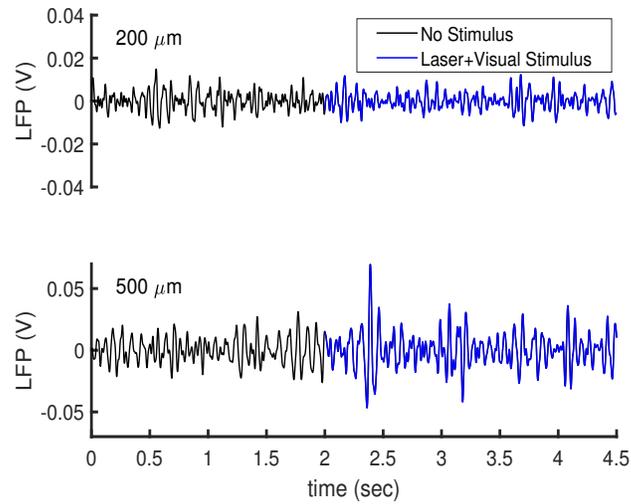}
\caption{LFP in Laser+Visual condition in different depths.}
\label{fig:f1}
\end{figure}

\begin{figure}
\centering
\includegraphics[width=8cm, height=6.5cm]{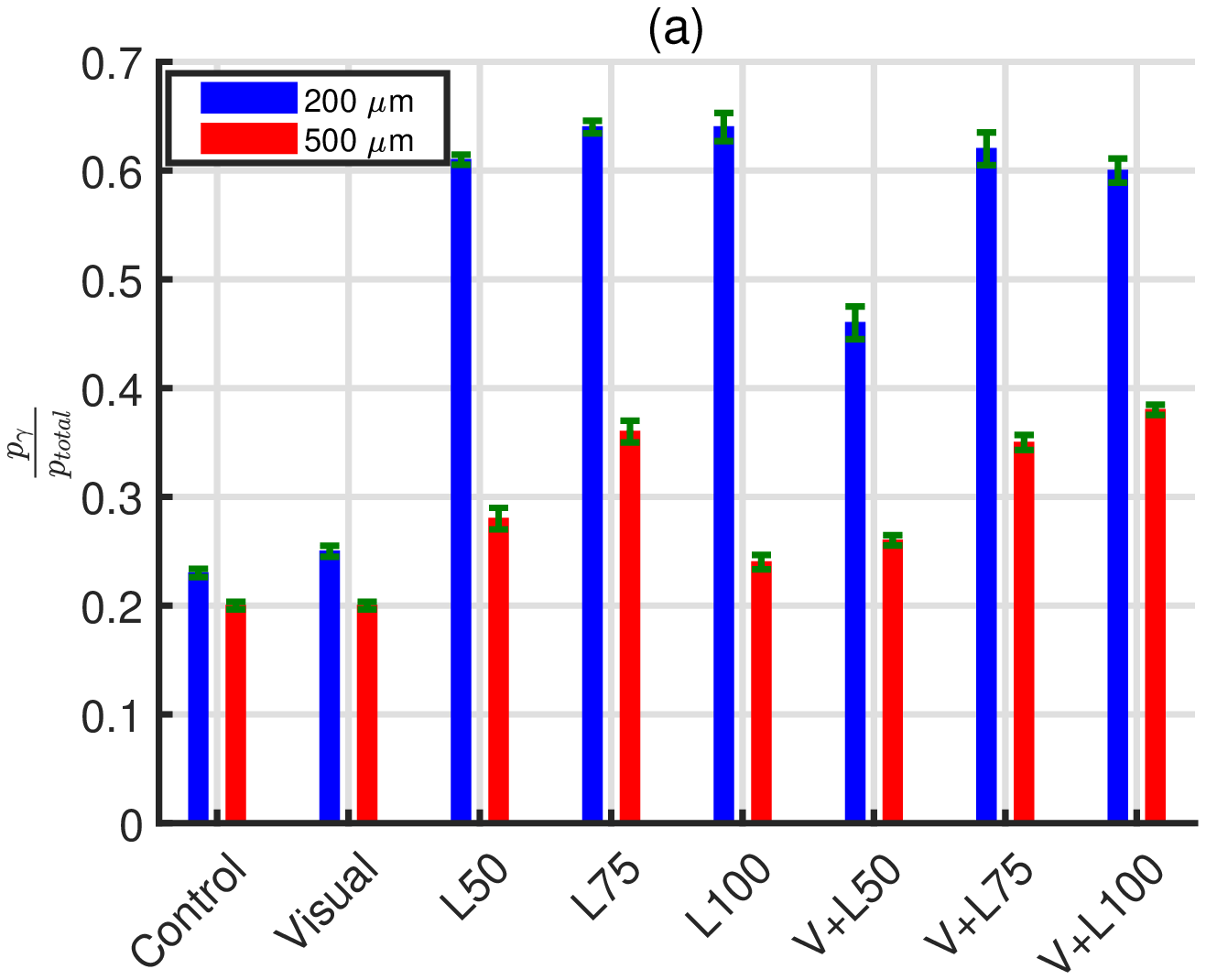}
\includegraphics[width=8cm, height=6.5cm]{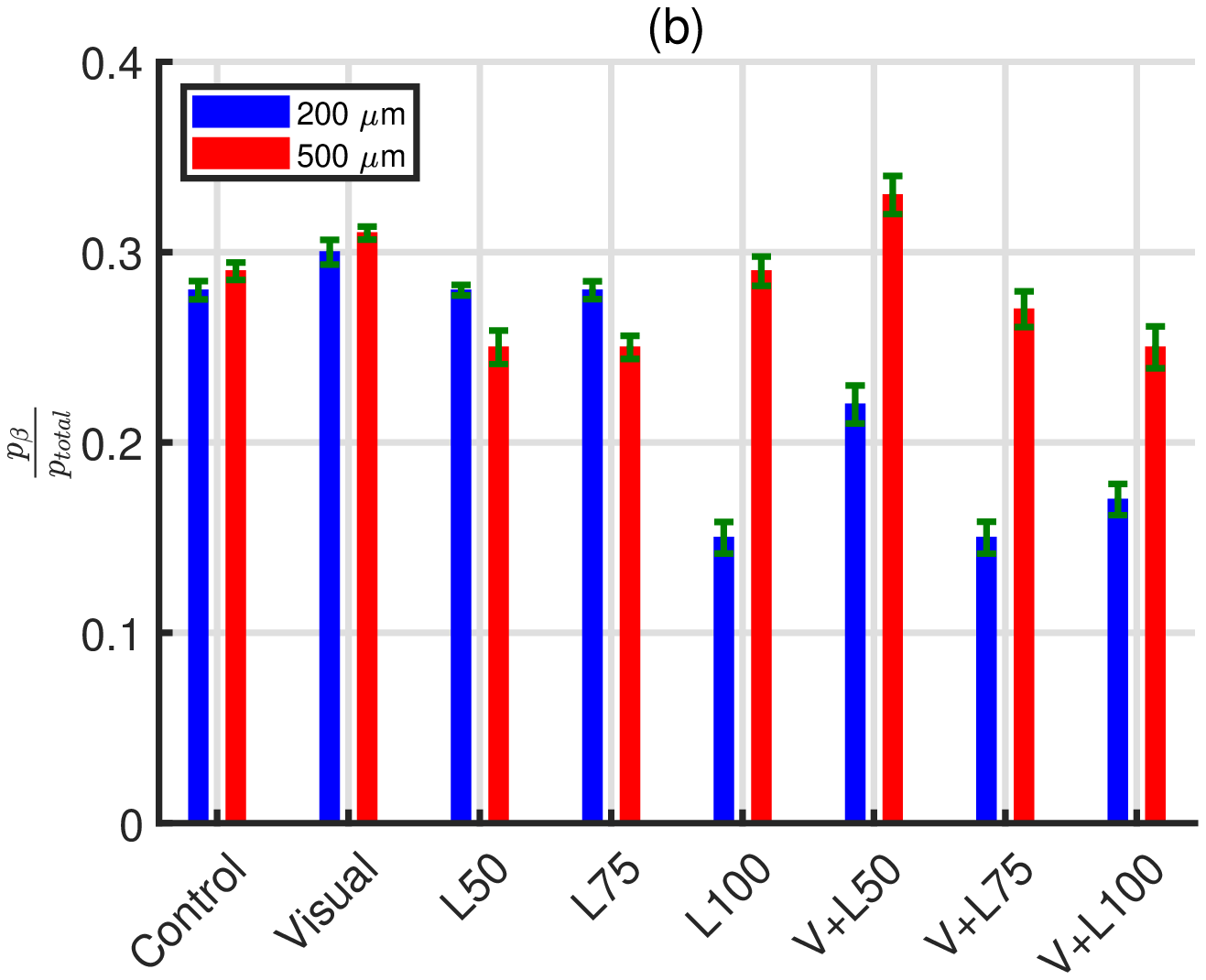}
\caption{ (Color online)  a: Mean Gamma power; b:Mean Beta power in Laser+Visual condition and different depths. Red for 500 $\mu$m and blue for 200 $\mu$m.}
\label{fig:f2}
\end{figure}

Figure~\ref{fig:f1} shows Local Field Potential (LFP) sample in Laser+Visual stimulus. In the two first seconds there is no laser stimulus, and it is switched-on after that. We first focus on variations in Gamma and Beta frequencies power in different conditions.\\
Figure \ref{fig:f2}a shows Gamma power in different conditions. As shown in figure \ref{fig:f2}, Gamma power increases in laser trials in both 200 and 500 $\mu$m, which means optogenetic stimulus induces more powerful Gamma oscillations in comparison with visual trials. However, optogenetic laser leads to Gamma oscillations; in 200 $\mu$m, Gamma power is more significant than 500 $\mu$m. When the subject is under both visual and laser stimuli, Gamma power decreases in both depths (except 500 $\mu$m with 100 mW/mm$^2$ intensity laser); it starts to increase when lasers' intensity increases.\\
Figure ~\ref{fig:f2}b shows Beta power in different conditions. When there is only laser stimulus, Beta power decreases in both depths. When the subject is under both visual and laser stimuli, Beta power decreases in 200 $\mu$m (except in 100 mW/mm$^2$ intensity laser). In 500 $\mu$m in Laser+Visual trials, the first Beta power increases and starts to decrease by implementing high laser intensity.
In order to see if optogenetic leads to have different conditions in vision we compare P$_{\text{Visual}}$ - P$_{\text{Control}}$ and P$_{\text{Laser+Visual}}$ - P$_{\text{Laser}}$ , we define it as biological response (see materials and methods) . The results for 500 $\mu$m and 200 $\mu$m are summarized in table~\ref{table:t1} and table~\ref{table:t2} respectively. Both tables show that the Optogenetic laser has a different influence on visual power in both depths. The laser's impact is mainly negative in 200 $\mu$m for both Beta and Gamma oscillations, while for laser intensities 50 mW/mm$^2$ and 100mW/mm$^2$ are positive for both Beta and Gamma oscillations at 500 $\mu$m.
\begin{table}[h!]
\centering
 \begin{tabular}{|c| c c c|} 
 \hline
 & P$_{\text{Visual}}$ - P$_{\text{Control}}$ & P$_{\text{Visual}+}$$_{\text{Laser}}$ - P$_{\text{Laser}}$ &( P$_{\text{Visual}+}$$_{\text{Laser}}$ - P$_{\text{Laser}}$)-(P$_{\text{Visual}}$ - P$_{\text{Control}}$ )  \\ [0.5ex] 
 \hline
 laser intensity & 0 & 50 \ \ \ \ \ \ \ \ \ \ \ \ \ \ \ \ \  75 \ \ \ \ \ \ \ \ \ \ \ \ \ \ \ \ \ 	 100 & 50   \ \ \ \ \ \ \ \ \ \ \ \ \ \ \ \ \  75  \ \ \ \ \ \ \ \ \ \ \ \ \ \ \ \ \ 	 100 \\ 
 mW/mm$^2$: &  &  \ \ \ \ \ \ \ \ \ \ \ \ \ \ \ \ \   \ \ \ \ \ \ \ \ \ \ \ \ \ \ \ \ \ 	  &   \ \ \ \ \ \ \ \ \ \ \ \ \ \ \ \ \    \ \ \ \ \ \ \ \ \ \ \ \ \ \ \ \ \ 	  \\ 
 
 \hline
 Frequency range:
 \\
 Beta  & 0.02$\pm$0.0087 &  0.08$\pm$0.0188  \ \ \ 0.02$\pm$0.0155  \ \ \ -0.04$\pm$0.0187 &  0.06$\pm$0.0275  \ \ \  $\pm$0.0242  \ \ \  - 0.06$\pm$0.0247 \\ \\

 Gamma  & $\pm$0.0069 &  -0.02$\pm$0.0147  \ \ \  -0.01$\pm$0.0169  \ \ \ 0.14$\pm$0.0114 &   -0.02$\pm$0.0216  \ \ \   -0.01$\pm$0.0238  \ \ \   +0.14$\pm$0.0183 \\[1ex] 

 \hline 
 \end{tabular}
\caption{ Biological response in 500 $\mu$m.}
\label{table:t1}
\end{table}
\begin{table}[h!]

\centering
 \begin{tabular}{|c| c c c |} 
 \hline 
 & P$_{\text{Visual}}$ - P$_{\text{Control}}$ & P$_{\text{Visual}+}$$_{\text{Laser}}$ - P$_{\text{Laser}}$ &  ( P$_{\text{Visual}+}$$_{\text{Laser}}$ - P$_{\text{Laser}}$)-(P$_{\text{Visual}}$ - P$_{\text{Control}}$ ) \\ [0.5ex] 
 
 laser intensity & 0 & 50  \ \ \ \ \ \ \ \ \ \ \ \ \ \ \ \ \  75  \ \ \ \ \ \ \ \ \ \ \ \ \ \ \ \ \ 	 100  & 50   \ \ \ \ \ \ \ \ \ \ \ \ \ \ \ \ \  75   \ \ \ \ \ \ \ \ \ \ \ \ \ \ \ \ \ 	 100   \\ 
 mW/mm$^2$: &  &  \ \ \ \ \ \ \ \ \ \ \ \ \ \ \ \ \   \ \ \ \ \ \ \ \ \ \ \ \ \ \ \ \ \ 	  &   \ \ \ \ \ \ \ \ \ \ \ \ \ \ \ \ \    \ \ \ \ \ \ \ \ \ \ \ \ \ \ \ \ \ 	  \\ 
 
 \hline
 
 Frequency range: 
\\
 Beta  & 0.02$\pm$0.0113 &  -0.06$\pm$0.0128  \ \ \  -0.13$\pm$0.0131  \ \ \  -0.04$\pm$0.024 & -0.08$\pm$0.0241  \ \ \  -0.15$\pm$0.0244   \ \ \  -0.06$\pm$0.0353 \\ \\

 Gamma & 0.02$\pm$0.0089 &  -0.25$\pm$0.0196   \ \ \  -0.02$\pm$0.0208  \ \ \   0.02$\pm$0.0165 &  -0.027$\pm$0.0285  \ \ \   -0.04$\pm$0.0297  \ \ \   $\pm$0.0254 \\[1ex] 

 \hline
 \end{tabular}
\caption{ Biological response in 200 $\mu$m.}
\label{table:t2}
\end{table}

\begin{figure}
\centering
\includegraphics[width=8cm, height=6cm]{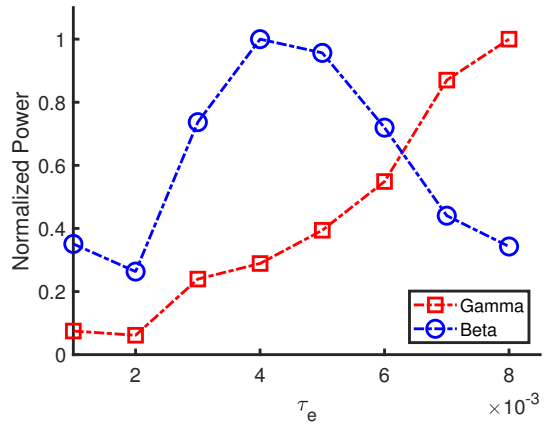}
\caption{ Gamma and Beta power versus $\tau_{e}$: monatomic behavior of Gamma power, and nonmonatomic behavior of Beta power by considering $\tau_{i}=0.004$.}
\label{fig:f3}
\end{figure}

Our approach to understanding the basic behind the results is Wilson-Cowen model \cite{heitmann2017optogenetic} .
The specific range of injected currents leads to Gamma oscillations in population activity \cite{onslow2014canonical}. Besides, Optogenetic stimulus active inhibitory and excitatory neurons by changing their firing rates \cite{heitmann2017optogenetic}. Because of opsin's type, the Optogenetic laser excites excitatory neurons; therefore, we consider a pulse input to the excitatory neurons; however, this current does not significantly change the Gamma power. 
 Figure ~\ref{fig:f3} shows the amount of Gamma power per power of the whole signal versus the different excitatory time constants.  Figure ~\ref{fig:f3} shows while excitatory time constat is decreasing, the power of Gamma is increasing. It can describe why the Gamma power in 200 and 500 $\mu$m increases in laser mode compared to visual and control modes. Therefore we conclude the Gamma power changes might arises due to the changes in population time constants.

We concluded in Laser+Visual states that vision acts as a resistance and prevents the excitatory populations from decreasing time constant. As shown in figure ~\ref{fig:f2}, Gamma power in Laser+Visual is less than in laser trials; however, it is still more than visual and control states.\\
As figure ~\ref{fig:f3} b shows, Gamma oscillations show a monotonic behavior; however, Beta oscillations are nonmonotonic (Figure ~\ref{fig:f3} ). Figure ~\ref{fig:f3}  shows the nonmonotonic behavior of Beta oscillations in both depths. 
\\
Therefore, our modeling approach suggests that Beta oscillations can increase or decrease depending on initial conditions; figure ~\ref{fig:f3} shows this nonmonotonic behavior in empirical results. As described before, Optogenetics is amplifying Gamma oscillations by decreasing the time constant of the excitatory population. If the initial time constat of excitatory cells was near the peak in figure ~\ref{fig:f3}, then the decreasing time constant leads to less Beta oscillation, which is in agreement with Figure ~\ref{fig:f3} in laser trials. However, in the laser+visula trials, while vision acts as a resistance in 500 $\mu$m, it is a resonator in 200 $\mu$m and decreases Beta oscillations significantly. One can say, vision tries to push time constant to peak in 500 $\mu$m while away from the maximum in 200 $\mu$m. \\

\section*{Discussion}

Gamma and Beta Oscillation's average power in control and visual trials at 500 $\mu$m is slightly more than 200 $\mu$m. However, Gamma oscillations are amplified at 200 $\mu$m, and Beta oscillations (except for laser trials at the intensities of 50 and 75 mW/mm$^2$) show similar behavior at 500 $\mu$m. Gamma oscillations at all trials at 200 $\mu$m have more power than 500 $\mu$m. Except for laser trials at intensities of 50 and 75 mW/mm$^2$, Beta oscillations power at 500 $\mu$m is slightly more than 200 $\mu$m.  Our modeling approach (see section "Modeling Approach") suggests that results at  50 and 75 mW/mm$^2$ in laser trials arise from Beta oscillations' nonmonotonic behavior versus the changing time constant of excitatory populations.

\subsection*{Layer II}

The Gamma power in all three laser intensities is higher than in the control mode, where our model describes this behavior by shifting the excitatory time constant to the higher amounts.  
Moreover, Gamma power in all three Laser+Visual trials is higher than the visual trials. Therefore, our results indicate a positive effect of the Optogenetics on the induction of Gamma oscillations.
The power of Gamma oscillation in laser trials is higher than in Laser+Visual trials. While our modeling approach suggests Optogenetics laser shift excitatory time constant to the higher amounts; it seems vision acts as resistance and prevents shifting a lot.
The Beta oscillations power in laser and Laser+Visual trails are less than control and visual trials, and Optogenetics laser is reducing the power in all trials. However, the Optogenetics laser reduces Beta power; Laser+Visual trials have less Beta power than laser trials. Our modeling approach suggests that it is due to Beta oscillations' nonmonotonic behavior, and we predict vision and Optogenetics laser corporate to reduce Beta oscillations power in 200$\mu$m. In contrast, as discussed above, vision makes it resistant to prevent sharp increases in Gamma oscillations.
A comparison of two laser and Laser+Visual trials indicates that the laser trials have more substantial beta oscillation at 50 and 75 mW/mm$^2$ intensities. However, at the 100 mW/mm$^2$ intensity, the Laser+Visual state's beta-oscillation is more than laser trials. The excessive intensity of the laser seems to reverse the behavior entirely.

\subsection*{Layer V}

The laser trials' Gamma power is higher than the control trials, and the Laser+Visual trials are higher than the visual trials. Therefor 
Optogenetic amplifies the power of Gamma oscillations at 500 $\mu$m.
By considering laser and Laser+Visual trials, vision prevents sharp increment in Gamma power; even with different laser intensities.
Beta oscillations power in laser trials at 50 and 75 mW/mm$^2$ compared to the control trials is reducing. However, both laser and control are approximately at the same level of Beta power in intensity. Our modeling approach (Figure 4) suggests that if the initial time constant of excitatory is near the peak, changing the time constant can decrease Beta power. Increasing Gamma power while decreasing Beta power is due to Gamma oscillations' monotonic behavior and nonmonotonic behavior of Beta oscillation.
\\
Here we observed two different vision behavior in 200 $\mu$m and 500 $\mu$m. While vision in cooperation of laser reduced the Beta power in 200 $\mu$m, it shows the opposite behavior in 500 $\mu$m. 
According to our modeling approach, it could be due to the different time constant of populations in different depths.
Generally, the Optogenetic laser amplifies Beta oscillations' power and reduces the Beta oscillations' power at both depths. 
Biological responses at 200 $\mu$m are negative in all cases, and at 500 $\mu$m, it is highly dependent on laser power. Laser intensity and visual power have an indirect relationship in Beta oscillations; while it is a direct relationship for Gamma oscillations, we have concluded cortical responses for Gamma and Beta oscillations are opposite. \\
In our experiment, the opsin affects the excitatory population, and then we consider a laser pulse current to the excitatory population. The different laser injection amplitudes can make the excitatory population faster or slower via changing the time constant. As the excitatory population gets fast, then the inhibitory population will get fast too, and therefore some oscillatory behaviors could emerge.\\
We conclude that when optogenetic stimulus changes the time constant of the inhibitory and excitatory neurons, specific amounts of injected current increase or decrease the frequencies power.

\subsection*{Conclusion}
\begin{figure}[hbt!]
\centering
\includegraphics[width=5.5cm, height=4.5cm]{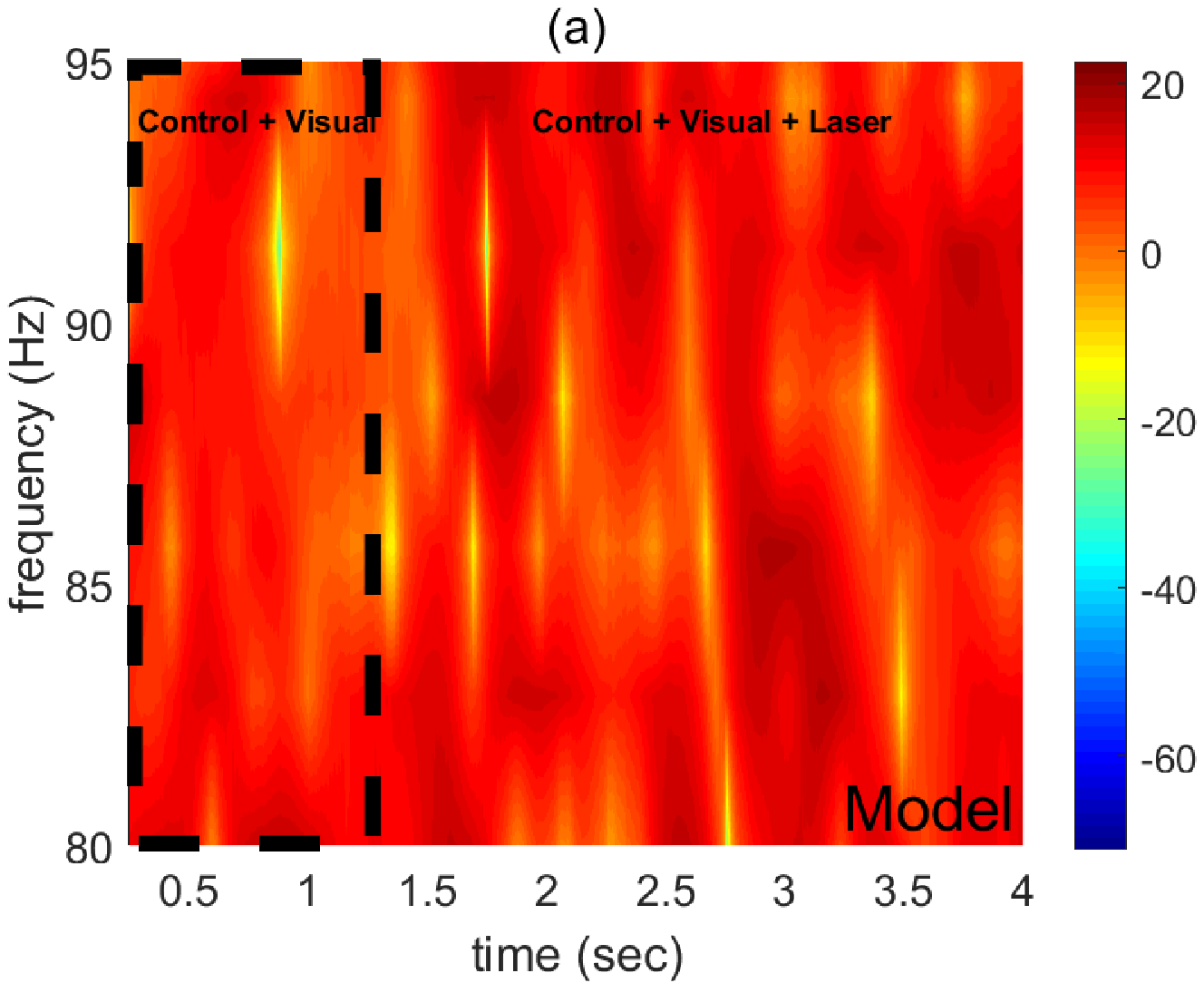}\quad
\includegraphics[width=5.5cm, height=4.5cm]{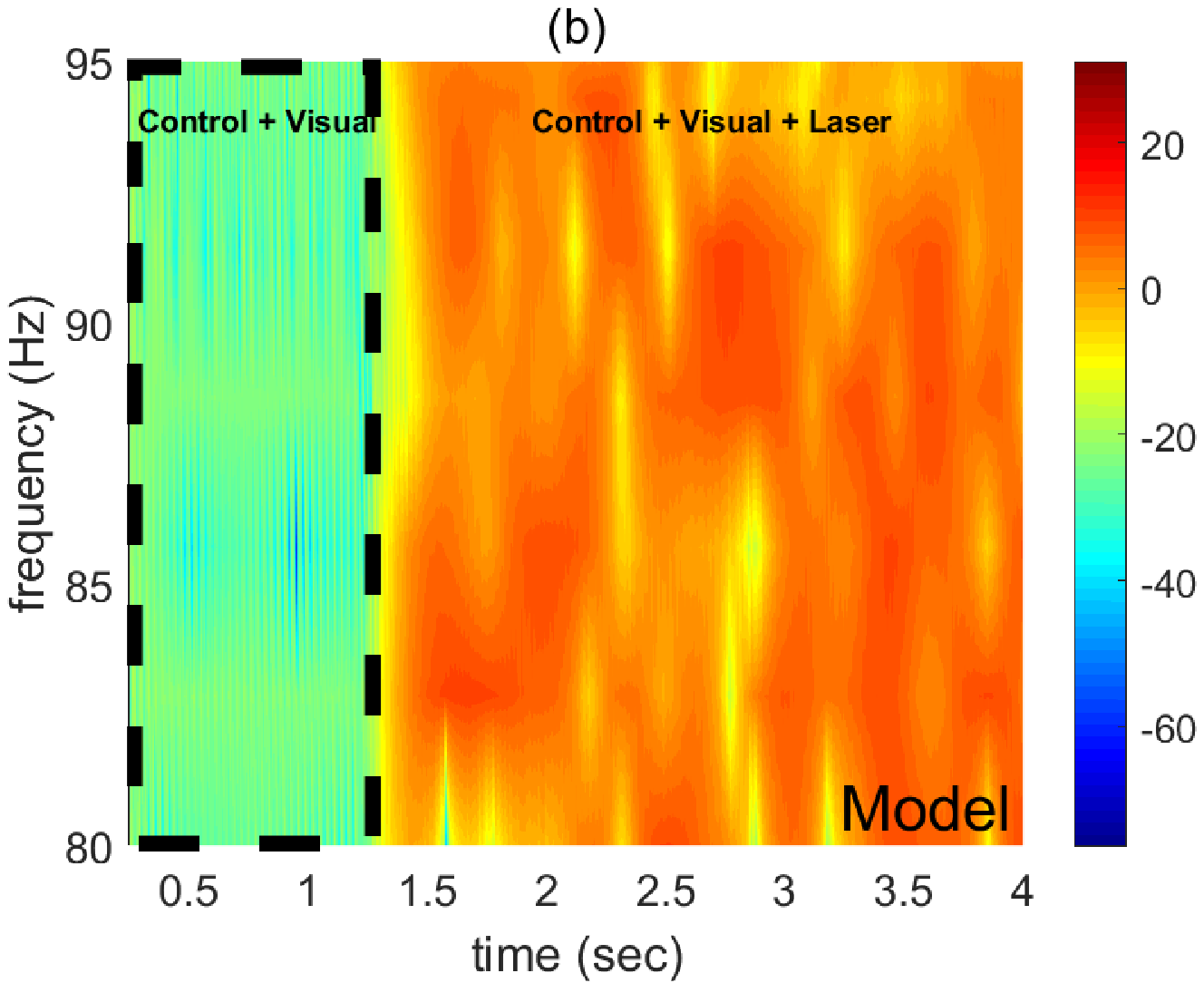}\\
\includegraphics[width=5.5cm, height=4.5cm]{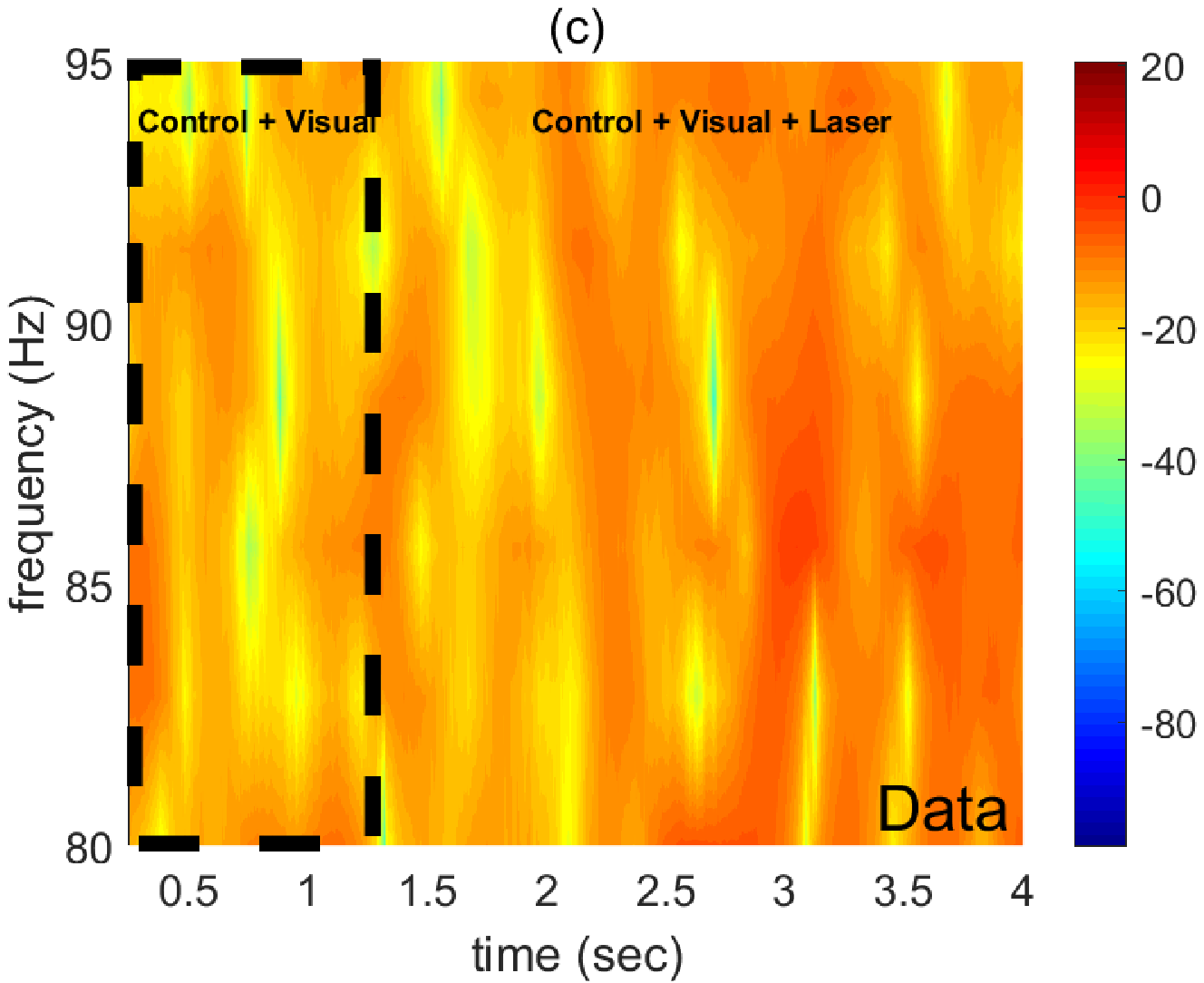}\quad
\includegraphics[width=5.5cm, height=4.5cm]{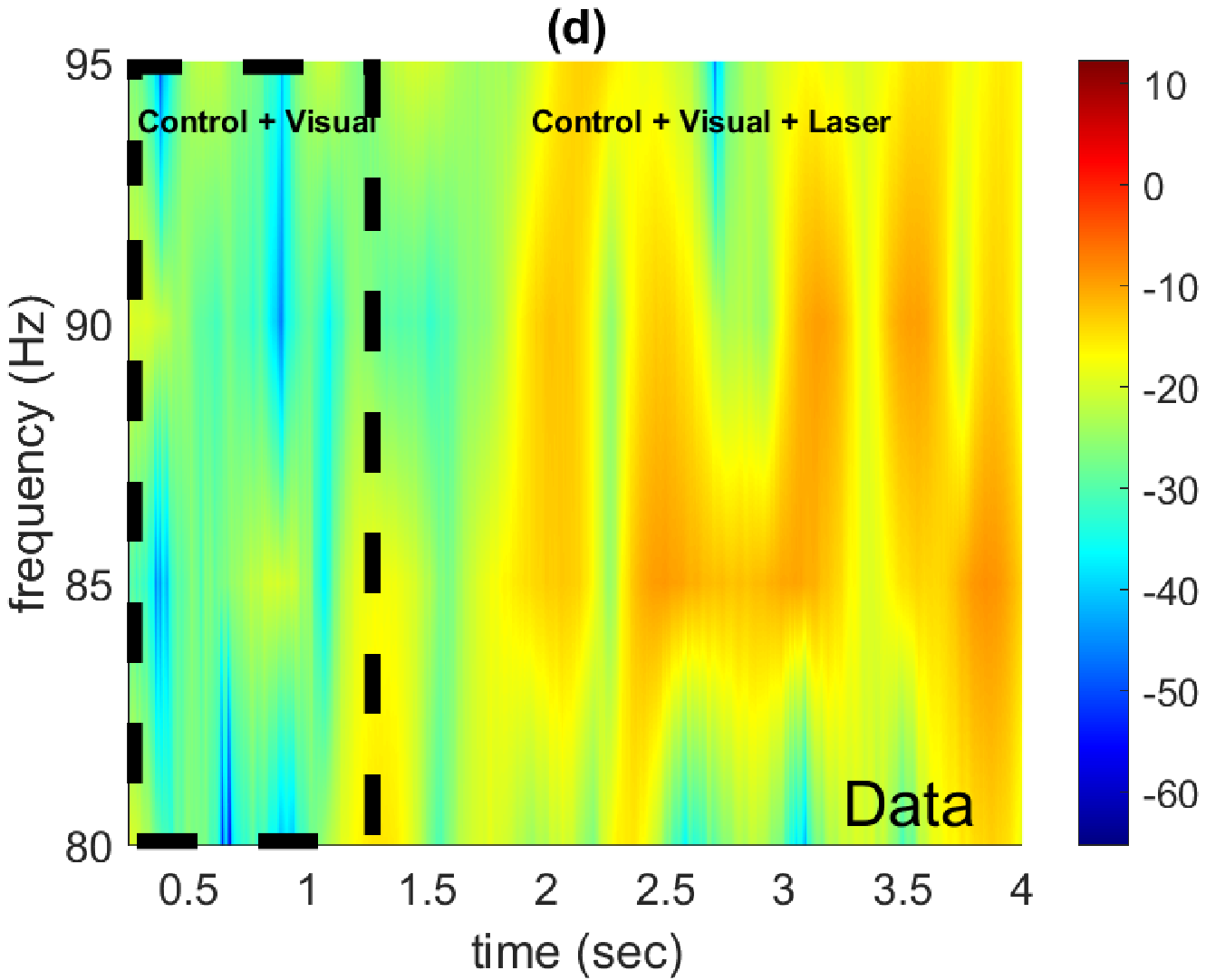}\\
\caption{ (Color online) Variation of Gamma  power in response to different amounts of excitatory and inhibitory time constants. a: $\tau_{e}$$<$$\tau_{i}$ b: $\tau_{e} $$>$$\tau_{i} $ c:  The fifth trial in Laser Visual  in 200 microns d: The Third trial in Laser+Visual in 500 microns.}
\label{fig:f7}
\end{figure}

Our simulations propose that considering different time constant of each population plays the most crucial role in modeling the stimulus\cite{heitmann2017optogenetic}.
The proposed model suggests that each population's time constant is changed under the optogenetic stimulus's influence, and the ratio of the time constants of the inhibitory and excitatory population explains the Optogenetic behavior.  The laser excites the excitatory cells, and it significantly changes Gamma power by changing the time constant of excitatory. Figure ~\ref{fig:f7} a shows no significant change in Gamma power in the whole time. There is no laser stimulus in the one and a half seconds, while there is a laser stimulus in the rest. Figure ~\ref{fig:f7}a agrees with Laser+Visual trial in 200 microns, Figure ~\ref{fig:f7}c. 
Figure ~\ref{fig:f7} b shows a significant change in Gamma power when the time constant of excitatory neurons are more than inhibitory neurons. 
Figure ~\ref{fig:f7} b can describe the data in figure ~\ref{fig:f7}d  which is a trial in 200 microns in Laser+Visual trial. 
As a conclusion, optogenetic stimulation in both depths of visual cortex induces gamma oscillations for all applied laser intensities, while beta oscillations are diminished for some intensities. In fact, the responses to Beta and Gamma oscillations are different; it is monotonic in Gamma oscillations but non-monotonic in Beta oscillations.\\

\section*{Acknowledgment}
All authors gratefully acknowledge financial support from their affiliated institution and thank Mehdi Aslani for graphical representation of the experimental setup. MDH is very thankful for valuable discussions with Udo Ernst for the theoretical modelling. \\
\bibliography{sample}
\end{document}